**Identification of suspicious behaviour through anomalies in the tracking data of fishing vessels**

**Jorge P. Rodríguez[1,*], Xabier Irigoien[2], Carlos M. Duarte[3], Víctor M. Eguíluz[4]**

1. Instituto Mediterráneo de Estudios Avanzados (IMEDEA), UIB-CSIC, Esporles, Spain
2. AZTI Marine Research, Pasaia, Spain
3. Red Sea Research Center (RSRC), and Computational Biosciences Research Center (CBRC), King Abdullah University of Science and Technology (KAUST), Thuwal, Kingdom of Saudi Arabia
4. Instituto de Física Interdisciplinar y Sistemas Complejos (IFISC), UIB-CSIC, Palma de Mallorca, Spain

**\*Corresponding author:**
jrodriguez@imedea.uib-csic.es
jorgeprodriguezg@gmail.com

**Keywords:** Automatic Identification System (AIS), fishing vessels, tracking data, suspicious behaviour, Exclusive Economic Zones (EEZ), Marine Protected Areas (MPA)

Automated positioning devices can generate large datasets with information on the movement of humans, animals and objects, revealing patterns of movement, hot spots and overlaps among others. This information is obtained after cleaning the data from errors of different natures. However, in the case of Automated Information Systems (AIS), attached to vessels, these errors can come from intentional manipulation of the electronic device. Thus, the analysis of anomalies can provide valuable information on suspicious behaviour. Here, we analyse anomalies of fishing vessel trajectories obtained with the Automatic Identification System. The map of silence anomalies, those occurring when positioning data is absent for more than 24 h, shows that they occur more likely closer to land, observing 94.9% of the anomalies at less than 100 km from the shore. This behaviour suggests the potential of identifying silence anomalies as a proxy for illegal activities. With the increasing availability of high-resolution positioning of vessels and the development of powerful statistical analytical tools, we provide hints on the automatic detection of illegal activities that may help optimise monitoring, control and surveillance measures.

**MAIN TEXT**

**Introduction**

Automatic Identification System (AIS) data provides valuable information about shipping activity in the oceans, revealing for example the appearance of new shipping routes (Eguíluz et al. 2016). In particular, concerning fishing vessels, global AIS analyses have provided inferences on the global distribution of fishing intensity (Kroodsma et al. 2018), the unequal share of the industrial fishing effort among countries (McCauley et al. 2018), the economic impact of fishing in the high seas (Sala et al. 2018), the network of fishing ports supporting fishing activity in different ocean regions (Rodríguez et al. 2021), or the overlap with marine animals (Queiroz et al. 2019).

Illegal, unreported and unregulated (IUU) fishing represents a problem for actors, both nations and companies, that cooperate to sustainably exploit fishing resources (Sumaila et al. 2006), leading for example to inaccuracies in the catch reports, hindering fisheries management. Furthermore, other illegal activities, such as forced labour, occur within the fishing industry (McDonald et al. 2021).

Interestingly, an initial screening of fishing vessels' AIS data (and in general any automated positioning system) reveals anomalous behaviours in the reported positions. For example, vessels often reported consecutive locations that would represent non-feasible trips according to typical vessel speeds, or the system omitted for a long time the otherwise frequently reported position, which we refer to as anomalous silence events. While the failure of the devices and processes involved in the transmission and reception of the data is a natural source of these anomalous positions, anomalies can also represent operational issues or intentional manipulation (Ford et al. 2018). Complementary datasets have been already used to show that vessels can intentionally disconnect the AIS system to stop reporting their locations while performing illegal activities at the sea. For example, data obtained from night light satellite pictures have reported illegal trajectories along North Korean waters (Park et al., 2020). Previous works have introduced statistical methods to detect anomalies or spoofed positions (Katsilieris et al. 2013). However, a global analysis of the distribution of AIS anomalies in fishing vessels that may provide cues to illegal or unreported activity is still pending. Here, we focus on silence anomalies, that is, long periods (compared to the typical AIS temporal resolution) without reporting AIS location. In particular, we aim to statistically identify geographical locations where fishing vessels show an excess of anomalous events that cannot be explained due to randomly distributed operational issues and are likely to represent intentional manipulation of these devices to hide the vessels' real position and trajectories.

**Results**

We identified 770K silence anomalies in our 2014 global fishing vessels movement dataset. For our analysis, we discretized the world in grid cells of size 0.5º lat x 0.5º lon and ignored grid cells which include harbours (see Methods and data), resulting in 25,795 grid cells with at least one observed anomaly, out of 87,853 total (non-harbour) grid cells with fishing vessels observations. Most anomalies occurred in grid cells with a high number of location estimates, such that these anomalies could be explained by random operational issues, happening more frequently in highly visited grid cells. Specifically, after randomising the location of each vessel's anomalies, such that they happen proportionally to the number of locations inside each grid cell observed in each trajectory, and selecting in each trajectory the grid cells with p-value < 0.01, 169K anomalies (22% of the anomaly observations) remained significant with respect to the random model (Figs. S1, S2, for a p-value < 0.05, 239K anomalies, i.e. 31% of the anomaly observations, remained significant). These significant anomalies, unlikely to be attributable to random events, were observed in 5758 unique grid cells (for p<0.05, 10577 unique grid cells), and appeared more frequently in the Northeastern Atlantic Ocean, the Mediterranean Sea and the Western Pacific (Eastern Asia shore) (Fig. 1a), generally implying low vessel velocities, irrespectively of the silence lag (Fig. S3). The distribution of the number of significant anomalies per grid cell displayed a heavy tail, indicating the presence of a few grid cells with very high numbers of anomalies (Fig. 1b). A key indicator of how frequent visits to grid cells ended up in anomalies was the fraction of anomalies, computed as the number of significant anomalies in each grid cell divided by the total number of locations from vessels in that grid cell, including only those vessels with at least one anomaly along their trajectories (Fig. 1c). This indicator revealed high fractions of anomalies in the Southwest of Ireland, included within the Sole Bank fishing ground, and the South of Japan (Fig. 1d).

Regarding the fishing vessels, 16K vessels displayed at least one significant anomaly, out of the 78K vessels included in our dataset. Although we observed at least one significant anomaly in only 20% of the analysed vessels, there were geographical hotspots where most of the detected fishing vessels displayed at least one significant anomaly (Fig. S4). Although this result illustrates the presence of hotspots with diverse fishing vessels showing significant anomalies, it does not inform about how often these vessels showed significant anomalies. In fact, our statistical filter captured both locations with many anomalies as preferential hotspots for AIS disconnection, but also locations where there were few visits of vessels with anomalous behaviour, but most visits lead to an event of silence (Fig. 2). Focusing on the regions where there were remarkable combinations of number and fraction of anomalies revealed spatial structures in the vicinity of the Exclusive Economic Zone (EEZ) limits of the United States, Ireland, Libya and Japan (Fig. 2a-h). In the region including the Bay of Biscay and the Atlantic coast of Ireland and Great Britain, the anomalies displayed a specific pattern following the limit of the West of Scotland Marine Protected Area (MPA) and appeared along the Bay of Biscay slope current (Fig. 2a,b). While the latter seems to be a geographical pattern linked to fishing pressure, the accumulation of anomalies at the border of the West of Scotland MPA provides high suspicion of illegal fishing activity. Another region of accumulation of anomalies was located close to the Northwest Atlantic shore (Fig. 2c,d), especially in the vicinity of the Northeast Canyons and Seamounts Marine Protected Area, again bordering an MPA. In the Mediterranean Sea, most of the anomalies were observed in the Italian EEZ, but interestingly the vessels that reached the border between Libyan and Greek EEZs displayed anomaly events in most of their visits (Fig. 2e,f), suggesting illegal fishing operations in these EEZs. The Northwest Pacific Ocean and the Sea of Japan displayed accumulations of anomalies at the border of the Russian EEZ and in the disputed control area adjacent to Japan, South and North Korea (Fig. 2g,h).

Considering only the distance from each grid cell to the coast, apart from other geographical features, we observed that the fraction of anomalies decreased with distance from the coast, a behaviour that we associate with the disconnection of the AIS system in the trips towards the ports, while this behaviour displayed an increase of more than 100% in the vicinity of the expected border of the EEZ limit at 200 nautical miles (Fig. 2i). This limit separates the Areas Beyond National Jurisdiction from the national waters, while the EEZ limits between two nations may be located at closer distances from the shore.

We connected each anomaly to the first visited port before and after its observation, such that if the vessel visited the port before and after, it accounted for one anomaly for that port, while anomalies with different origin and destination ports accounted for 0.5 anomalies for each port. 68.2% of the anomalies had the same origin and destination port while, at vessel level, 1) 47.5% of the vessels displayed all their anomalies from and to the same port, i.e. with sequences such as A-anomalies-A, B-anomalies-B and, within them, 96.8% of the vessels travelled always from/to a unique port; 2) 31% of the vessels showing anomalies either with common or different origins and destinations, that is, leading to sequences such as A-anomalies-B, but also A-anomalies-A or B-anomalies-B; and 3) 22% of the vessels with all their anomalies happening in the transit between two different ports. The distribution of anomalies among origin/destination ports was highly heterogeneous, with 164 ports (15% of the ports with at least 0.5 anomalies) associated with 85% of the anomaly-port trips. In particular, the top-20 ports, linked to 39.5% of the anomalies, included 14 ports in China, 2 in Spain, 2 in Italy, 1 in Ireland and 1 in the Netherlands (Fig. 3a). The distribution was broader when the anomaly-port links were reported by country, with 89% of the trips associated with only 16 countries (11% of the countries with at least 0.5 anomalies), and a top-20 list of countries which ports support fishing vessels contributing 91% of those anomalous trips (Fig. 3b).

The structure connecting anomalies to countries can be better understood when all the grid cells are associated with one nation, either its EEZ for grid cells located in Areas Under National Jurisdiction or the nearest EEZ for those located on the high seas (Figs. S5, S6). This classification allowed us to create a network connecting the locations of the anomalies to the countries whose ports were visited before and after the observation of those anomalies (Fig. 4). This network revealed a complex structure splitting into three regional groups (France-Spain-United Kingdom-Ireland, Norway-Iceland-United Kingdom, United States-Canada) and two nations with strong interactions within their own EEZs (China and Italy). After identifying the MMSI numbers from our database in the Vessel identity database of Global Fishing Watch (Kroodsma et al., 2018), we assessed the role of the vessel flags on this network, revealing frequently the same EEZ, flag and country where the supporting port is, but also interesting links, for example, vessels with Chinese flag with anomalies in the Argentinian EEZ, receiving support from ports in Chile and Uruguay (Fig. S7).

**Discussion**

Tracking data of fishing vessels has advanced our understanding of when and where fishing activities occur, facilitating the quantitative estimation of the captures thanks to modern algorithms, for example, those using Artificial Intelligence for measuring fishing effort from the vessels' trajectories. However, intentional manipulation of the devices that report the fishing vessels' locations may bias these analyses, providing underestimated results, especially in the locations where low or even null fishing effort is expected due to their conservation status. The availability of alternative datasets may help contrast these trajectories and reveal behaviours that AIS tracking data does not highlight, such as those reported thanks to the night lights from vessels (Geronimo et al. 2018, Park et al. 2020). Nevertheless, there may be other data gaps, such as those related to the resolution of satellite imagery or the absence of lights due to the cloud cover. Such problems highlight, as a preliminary step before contrasting different datasets, the use of new algorithms that detect anomalies within a single dataset (in parallelism with unsupervised learning), which later additional datasets can validate.

Identifying anomalies from automatically retrieved vessel positions has applications beyond fishing vessels and AIS in marine shipping. In general, anomaly detection refers to the identification of entries in a dataset that do not conform to the expected value and different techniques and methods have been developed (Chandola et al. 2009). Social data for example is affected by biases, inaccuracies (e.g. at the source of the data, processing), and methodological limitations (Olteanu et al. 2019). For example, the analysis of human mobility can be hindered by missing and extraneous locations, for instance, 75% of check-in traces are extraneous and can be associated with the social rewarding system (Zhang et al. 2013).

We have developed a method for extracting the locations where there is a statistically relevant accumulation of silence anomalies in fishing vessels' trajectories, as the combination of a high number of anomalies from specific vessels' trajectories with a low probability (< 1%) of occurring by chance. Using this approach, we detected the specific areas where these anomalies happened. We found four general cases that accounted for a large fraction of the oceanic hot spots of AIS anomalies, including (1) marine protected areas (Fig. 2a,c), (2) the edge of EEZs supporting productive fisheries (Fig. 2g)), (3) ocean areas disputed by different nations; for instance, the maritime border between Greece and Libya, which is known for experiencing territorial disputes, exhibits a low number of anomalies, but with a specific set of vessels that displayed an anomaly in most of their visits to that area (Fig. 2f); and (4) An additional case refers to anomalies by vessels within their own EEZ's, as exemplified by China and Italy. We hypothesise that this behaviour may be due to fishers willing to hide their chosen location from possible competitors, a behaviour deeply rooted in fishing culture. Fishermen are known not to share information freely and tend towards a culture of secrecy and deceit about the location of productive fishing sites (Evans et al. 2014), which is clearly at odds with the advent of AIS, where their position can be viewed by their competitors. Cases 1 and 2 above clearly point to illegal practices, suggesting that the approach developed here can effectively identify and monitor potential illegal fishing practices.

We observed quantitative and qualitative differences after classifying the vessels into short (length < 40 m) and long (length ≥ 40 m) fishing vessels. For example, most of the anomalies at the Great Sole Bank corresponded to short vessels, while the opposite behaviour occurred in the South of Japan (Fig. S8). Additionally, short vessels displayed their anomalies mainly in the proximity to the shore, as expected, but we observed that long vessels showed fewer anomalies close to the coast, and a remarkable increase in the proximity of the EEZ limits, indicating that the suspicious behaviour related to the borders of the high seas was mostly associated with longer vessels (Fig. S9).

Advances in AIS instrumentation and reporting will help remove anomalies caused by technical malfunction and possibly help report attempts at tampering with AIS. Our approach also identifies the main harbours where vessels that report suspicious anomalies use, which are not randomly distributed but are distributed in a limited subset of harbours within specific nations, pointing at geographical targets for inspections. We suggest using game theory to design the right incentives to encourage the more reluctant fishermen to use AIS reliably, as punitive measures alone will not be able to improve current levels of anomalies.

In summary, our results provide an approach to single out excess anomalies in AIS positions of vessels that may represent deliberate manipulation of the reported position for reasons that may extend from maintaining confidentiality on productive fishing grounds from competitors to illegal fishing activity. We also track the vessels to the harbours that support their operations and show that a small number of vessels support the fishing vessels reporting most of the anomalies. This analysis, based on a complex systems analysis of port-fishing vessel networks, extends the uses of AIS to further provide a tool to combat illegal fishing and conserve fish stocks.

**Materials and methods**

**Automatic Identification System (AIS) data.** The AIS system reports, among other information, the location of the vessels carrying it with high temporal resolution. We use AIS data of vessels categorized as 'fishing vessels' in 2014, with a total of $2.44 \times 10^8$ locations from 112,535 unique vessels, according to their Maritime Mobile Service Identity (MMSI) numbers.

**Presence of vessels in ports.** We obtained the global coordinates of ports from the World Port Index (2019) and considered a visit of any vessel to a port when the vessel was located in the grid cell (0.5º lat x 0.5º lon) assigned to that port.

**Detecting anomalous behaviour of silence.** We discarded the trajectory locations considered as visits to ports (see Presence of vessels in ports). Then, along the remaining trajectories, we inferred anomalous silence events when the time difference between two consecutively reported locations was longer than 24 h.

**Connecting anomalies to harbours.** We connected each anomaly to the first/last ports visited (see Presence of vessels in ports) after/before the detection of the anomaly.

**Statistical filter to remove random operational anomalies.** Given the total number of anomalies $N_{i,j}$ and the total number $D_{i,j}$ of data points from vessel $j$ at grid cell (0.5º lat x 0.5º lon) $i$ (excluding those assigned to ports), we distributed the total number of anomalies randomly among the grid cells, with probability proportional to $D_{i,j}$, obtaining $R_{i,j}^k$, the number of randomly located anomalies of vessel $j$ in grid cell $i$ for each realization $k$ (note that $\Sigma_i N_{i,j} = \Sigma_i R_{i,j}^k$ for any $k$). Then, we considered the p-value as the fraction of realizations where a grid cell had a higher number of anomalies in the random case than in the observed data:

$$p_{i,j} = \Sigma_k H(R_{i,j}^k - N_i)/J$$

where H(x) is the Heaviside function (0 if x is lower than 0 and 1 otherwise), and J is the number of realizations. We considered that the random distribution of anomalies could not explain the cells with p-value lower than 0.01 (p<0.05 results included in Supplementary Information), after generating $J=10^3$ independent realizations. Hence, the cells with p < 0.01 displayed a systematic excess of anomalies that could not be explained by a random distribution.

**Exclusive Economic Zones.** We extracted the Exclusive Economic Zones from MarineRegions and assigned to each 0.5º lon x 0.5º lat grid cell the EEZ that covers most of its area. However, the Areas Beyond National Jurisdiction, where intentional manipulation of AIS tracking devices is highly expected, were not included in this dataset. Hence, we assigned each grid cell in the high seas to the closest EEZ, according to a random walk in the grid where each cell was connected to its closest 8 neighbours (East, West, North, South, Northeast, Northwest, Southeast, Southwest, Fig. S5).

**Acknowledgments**

J.P.R. is supported by Juan de la Cierva Formación program (Ref. FJC2019-040622-I) funded by MCIN/AEI/ 10.13039/501100011033.

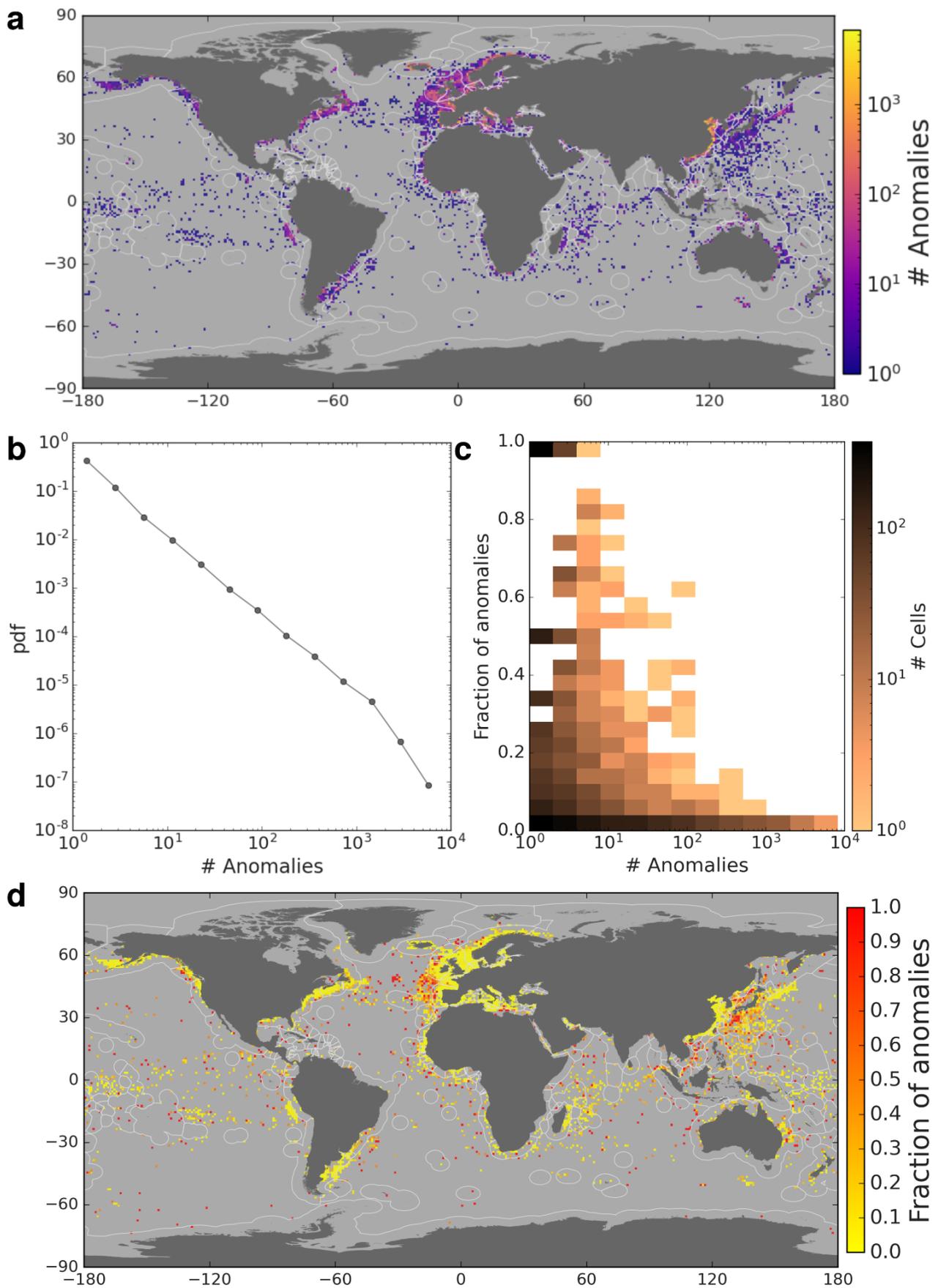

**Fig. 1. A global dataset of silence anomalies observations. a,** Geographical extent of the number of significant anomalies starting at each 1° lat x 1° lon grid cell. **b,** Distribution of the number of anomalies observed in each grid cell. **c,** Density plot of the fraction of significant anomalies, computed as the sum of significant anomalies in each grid cell divided by the number of locations of vessels with at least one anomaly in that grid cell, as a function of the total number of anomalies. **d,** Geographical extent of the fraction of anomalous locations. White contours on the maps represent the limits of Exclusive Economic Zones.

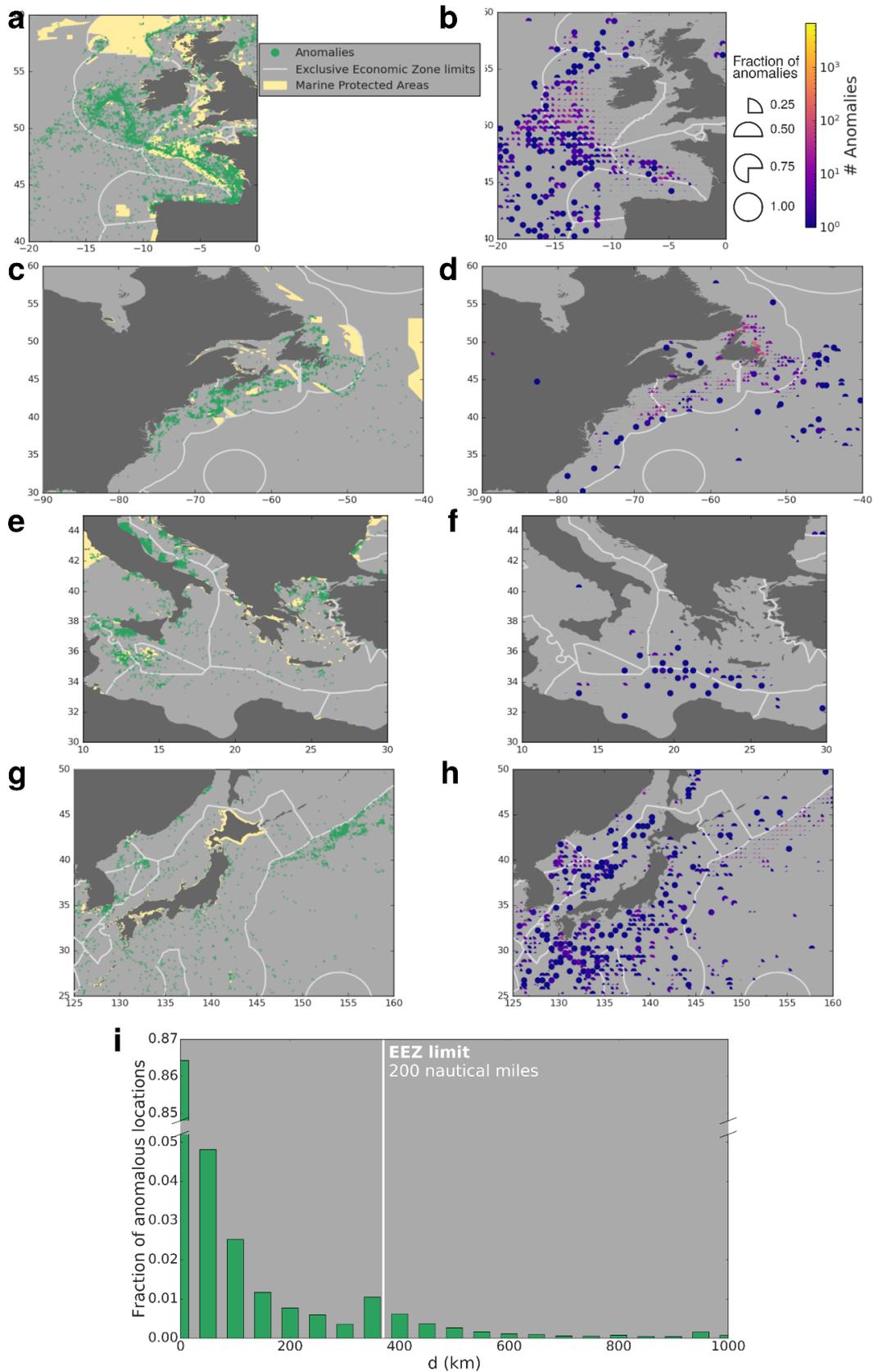

**Fig. 2.** Significant anomalous locations in the proximity of Exclusive Economic Zones (EEZ) and Marine Protected Areas (MPA). **a-h**, Regional distribution of anomalous locations. Panels **a,c,e,g** depict the location of the anomalies, while **b,d,f,h** describe the number (color) and fraction (fraction of the pie) of observed silence anomalies in each 0.5º x 0.5º grid cell. Regions correspond to Northeast Atlantic Ocean (**a,b**), Northwest Atlantic Ocean (**c,d**), Southern Mediterranean Sea (**e,f**) and Western Pacific Ocean (**g,h**). White contours and light yellow areas represent, respectively, the limits of the EEZ and the MPA. **i**, Fraction of the global number of anomalies observed in cells located at a distance d from the shore.

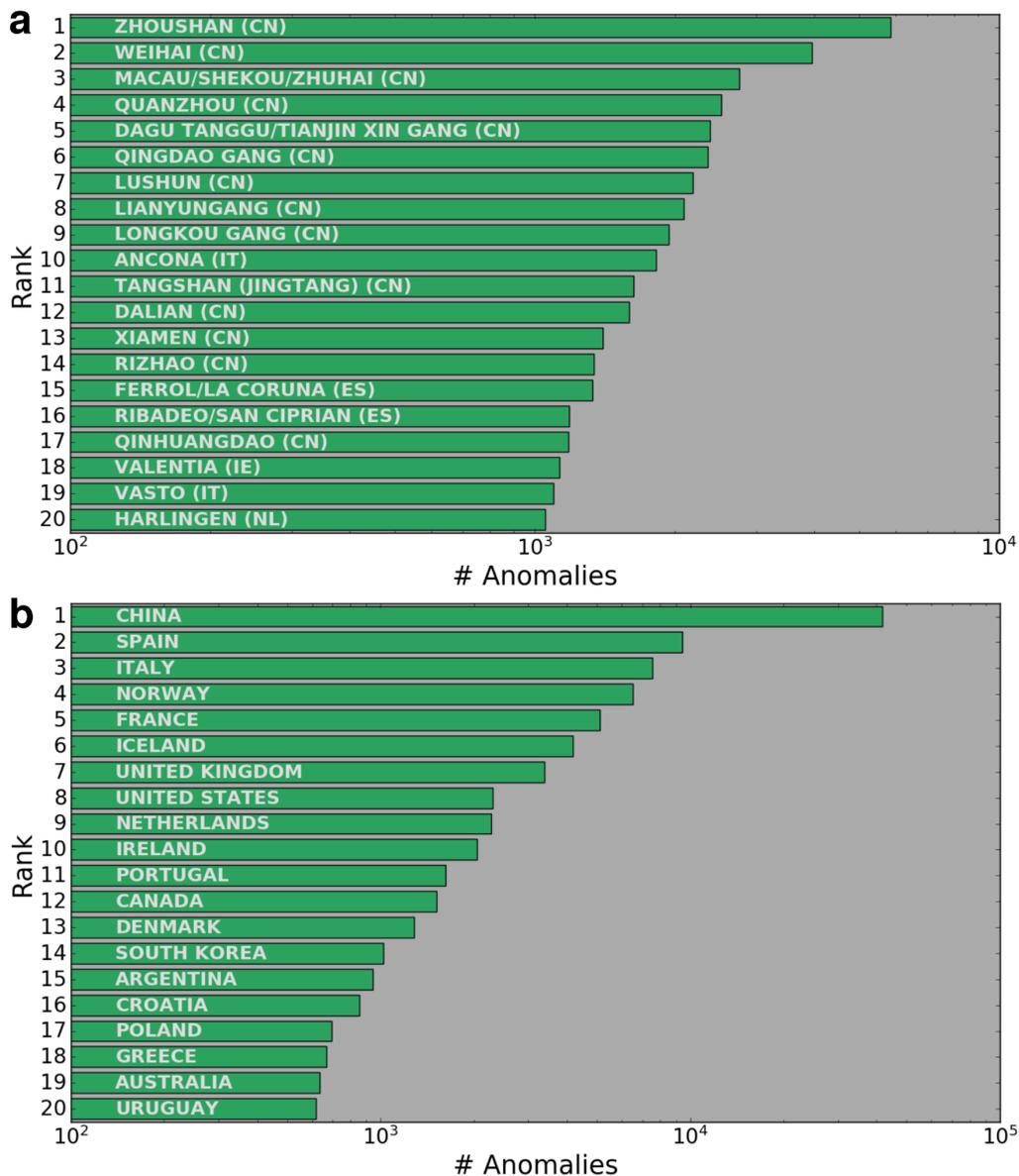

**Fig. 3.** Locations of the ports with anomaly detections before or after visiting them. **a**, Rank of the 20 ports with highest anomaly detections before or after visiting them. Countries are included within brackets. CN: China, IT: Italy, ES: Spain, IE: Ireland, NL: Netherlands. **b**, Rank of the 20 countries whose ports are visited before or after the anomalies.

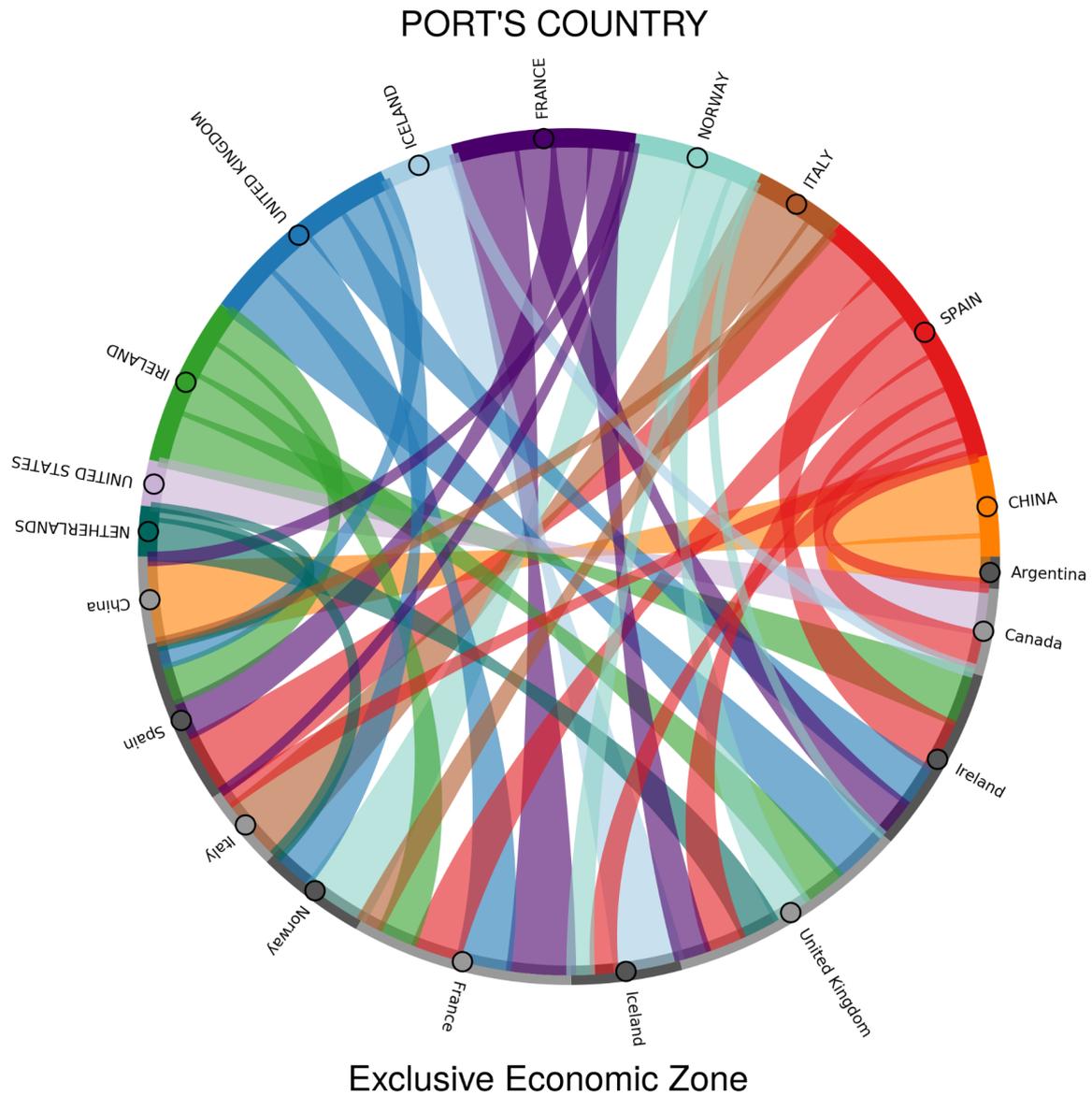

**Fig. 4.** Network between the top 10 countries (top, names in capital letters, colored nodes) with support from their ports to most of the observed anomalies, and the top 10 EEZs (bottom, full names, grey nodes) where the anomalies were observed. Link widths are proportional (logarithmic scale) to the number of anomalies connecting the nation's ports to anomalies in EEZs.

**SUPPLEMENTARY INFORMATION**

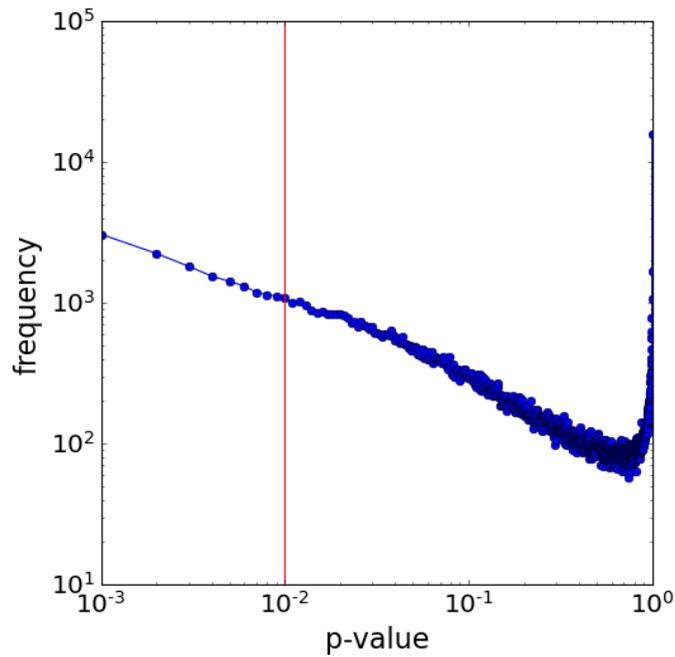

**Fig. S1.** Distribution of p-values. The red vertical line indicates the filter p<0.01 that we have used.

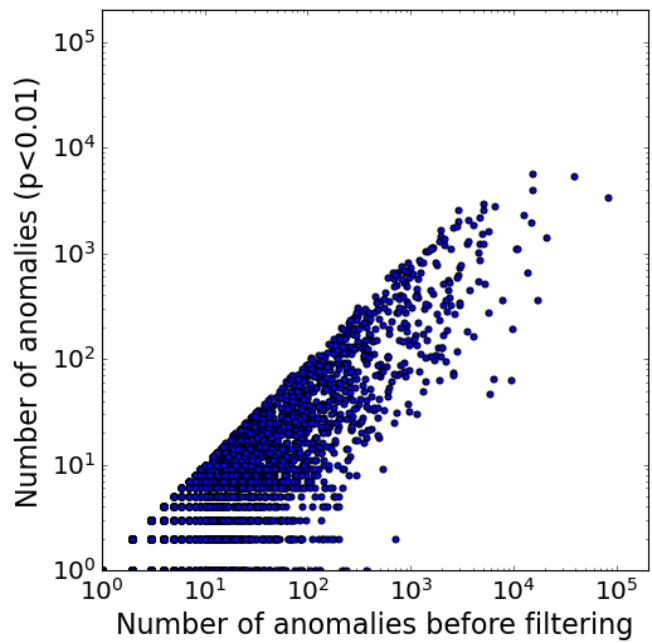

**Fig. S2.** Scatter plot of the remaining number of anomalies as a function of the original number of anomalies for the grid cells that keep at least one anomaly after filtering imposing p<0.01.

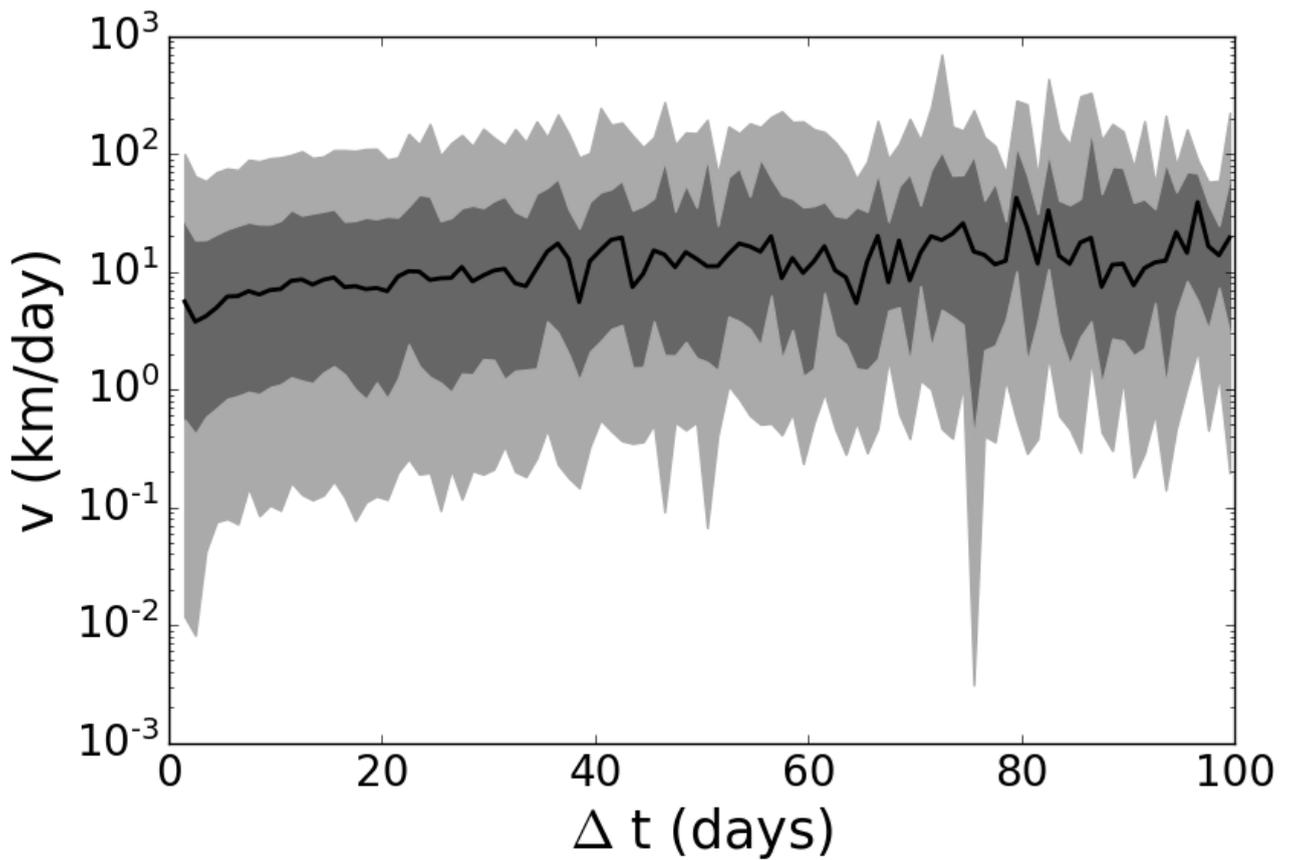

**Fig. S3.** Velocity measured on the events of silence anomalies. We measured the distance between the beginning and the end of the anomaly that had a silence lag of $\Delta t$, obtaining the vessel's velocity associated to that anomaly. The black curve represents the median, while the shaded areas represent the 25-75 percentiles (darker) and the 10-90 percentiles (lighter)

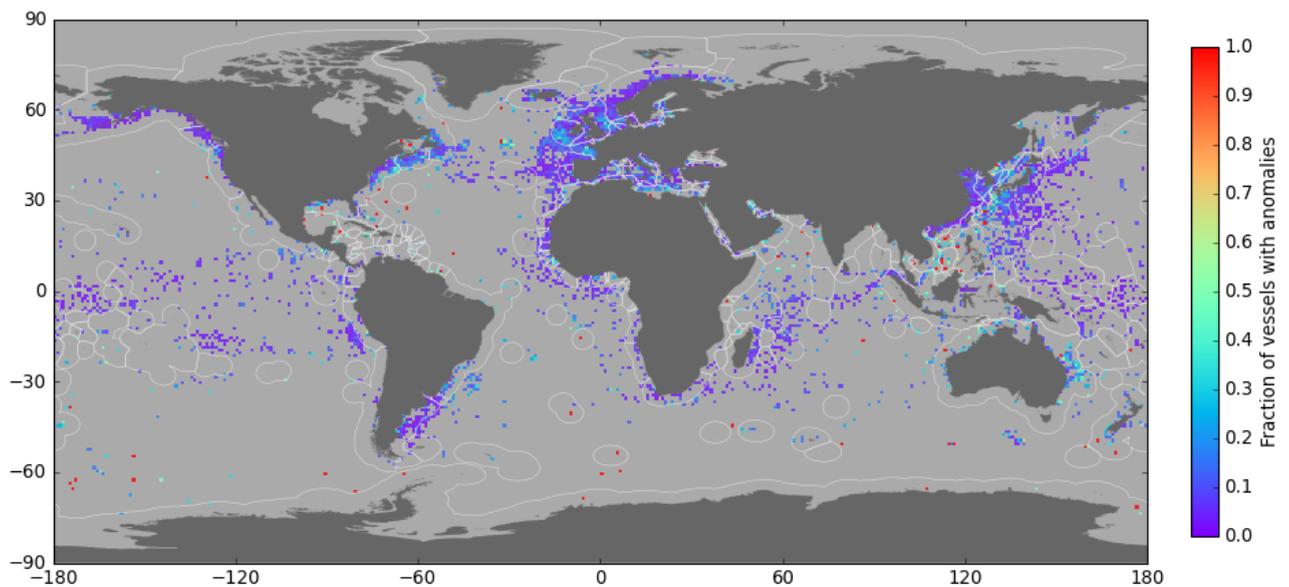

**Fig. S4.** Fraction of the detected vessels at each 1° x 1° grid cell displaying at least one significant anomaly within that cell.

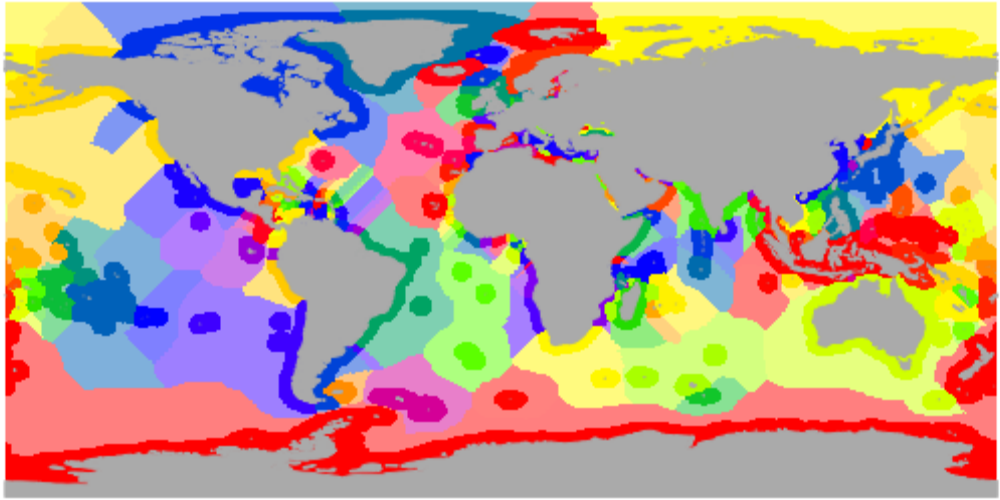

**Fig. S5.** Exclusive Economic Zones (bright colours) and their closest areas in the Areas Beyond National Jurisdiction (pale colours).

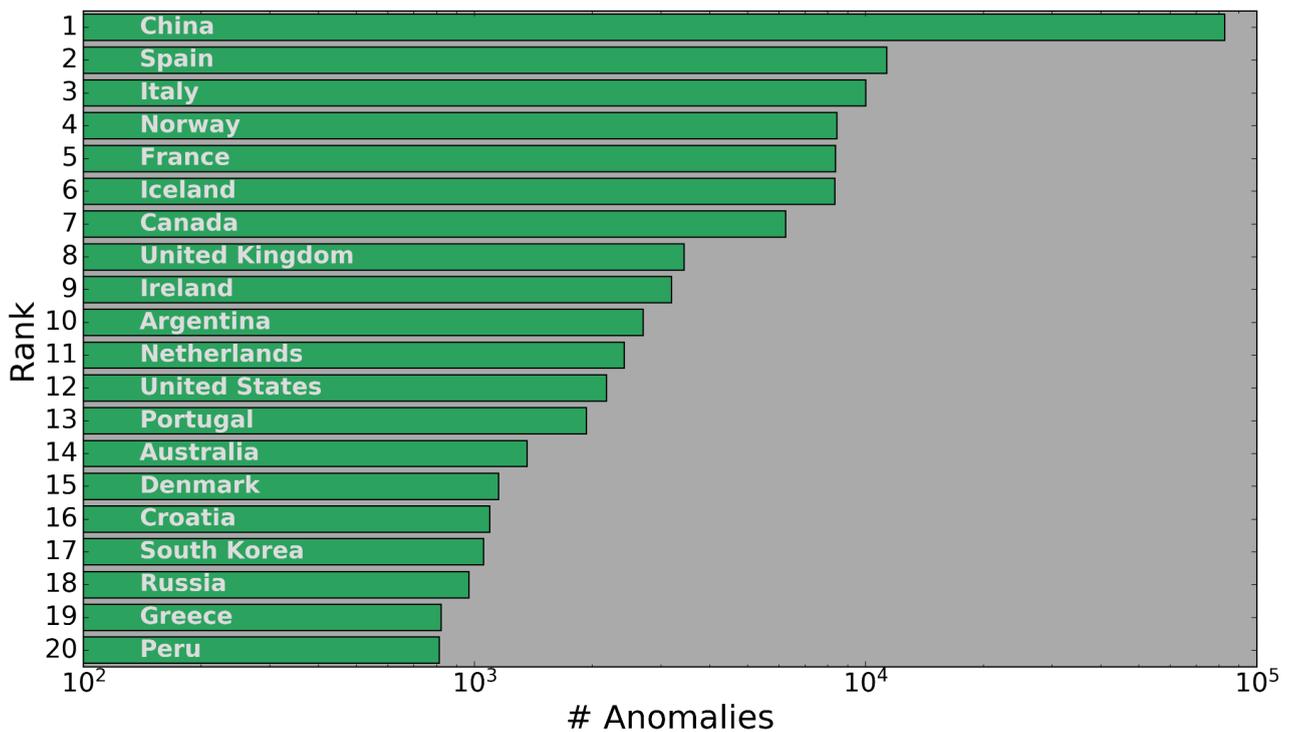

**Fig. S6.** Rank of the Exclusive Economic Zones (EEZ) with most of the anomalies associated with them. The closer cells to an EEZ than to any other are assigned to that EEZ (Fig. S1). The 20 EEZs represented here are associated with 94% of the observed significant anomalies.

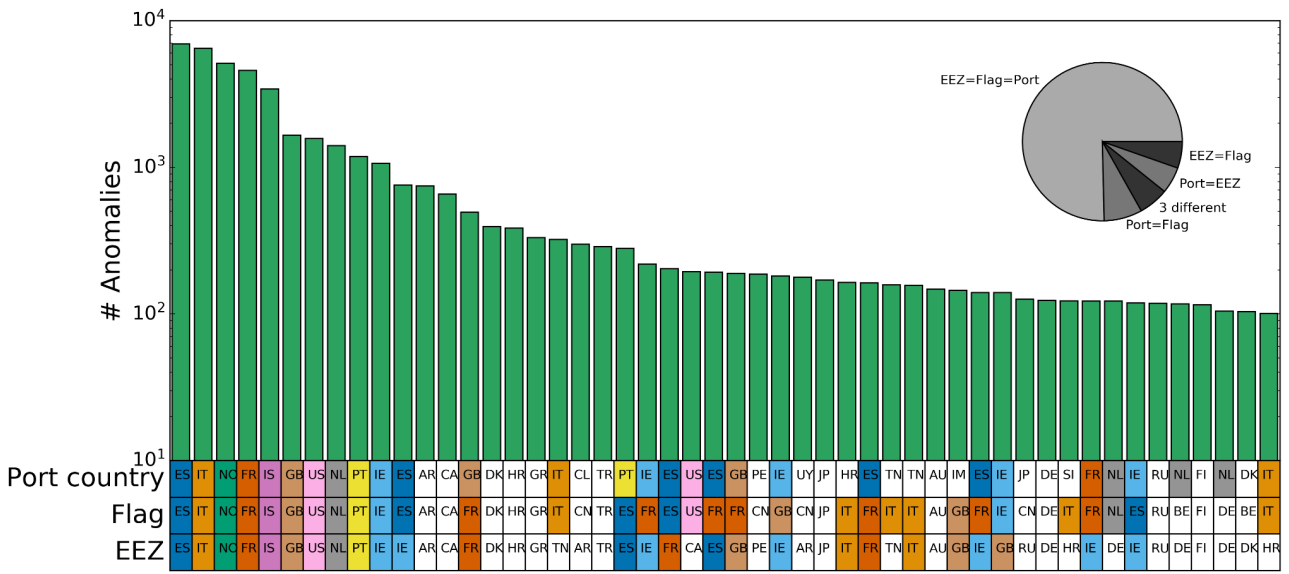

**Fig. S7.** Number of anomalies observed for each unique combination (Port country, Vessel flag, EEZ where the anomaly occurred). The anomalies account as 0.5 for the country before and after the anomaly, observed for the subset of MMSIs with available flag registry information (68 K anomalies). Colored cells correspond to the countries represented in the top-10 triplets. Country codes are ES: Spain, IT: Italy, NO: Norway, FR: France, IS: Iceland, GB: United Kingdom, US: United States, NL: Netherlands, PT: Portugal, IE: Ireland, AR: Argentina, CA: Canada, DK: Denmark, HR: Croatia, GR: Greece, TN: Tunisia, CL: Chile, CN: China, TR: Turkey, PE: Peru, UY: Uruguay, JP: Japan, AU: Australia, IM: Isle of Man, RU: Russia, DE: Germany, SI: Slovenia, BE: Belgium, FI: Finland.

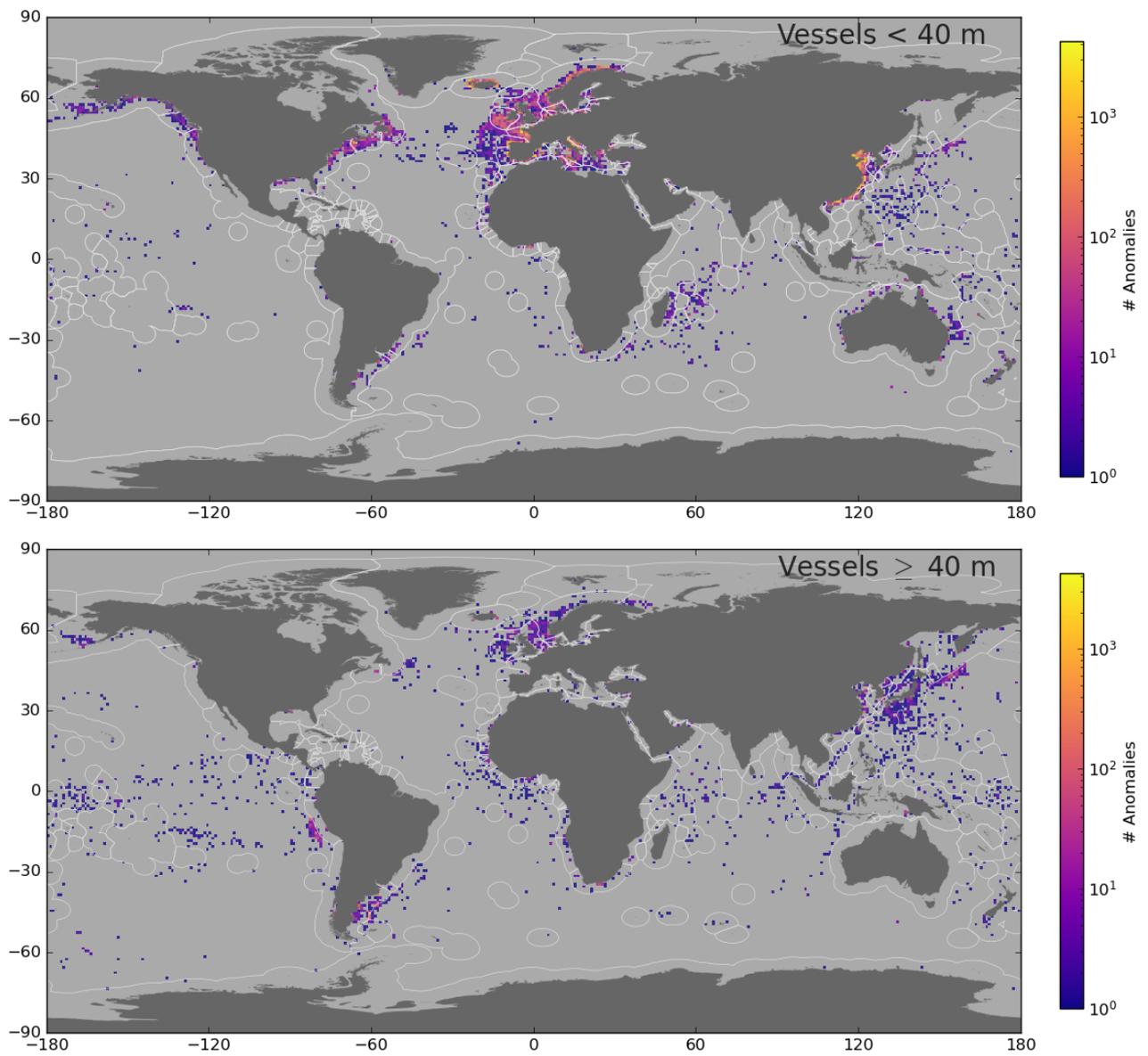

**Fig. S8.** Geographical extent of the number of significant anomalies starting at each 1º lat x 1º lon grid cell, for short (top, length < 40 m) and long (bottom, length ≥40 m) fishing vessels.

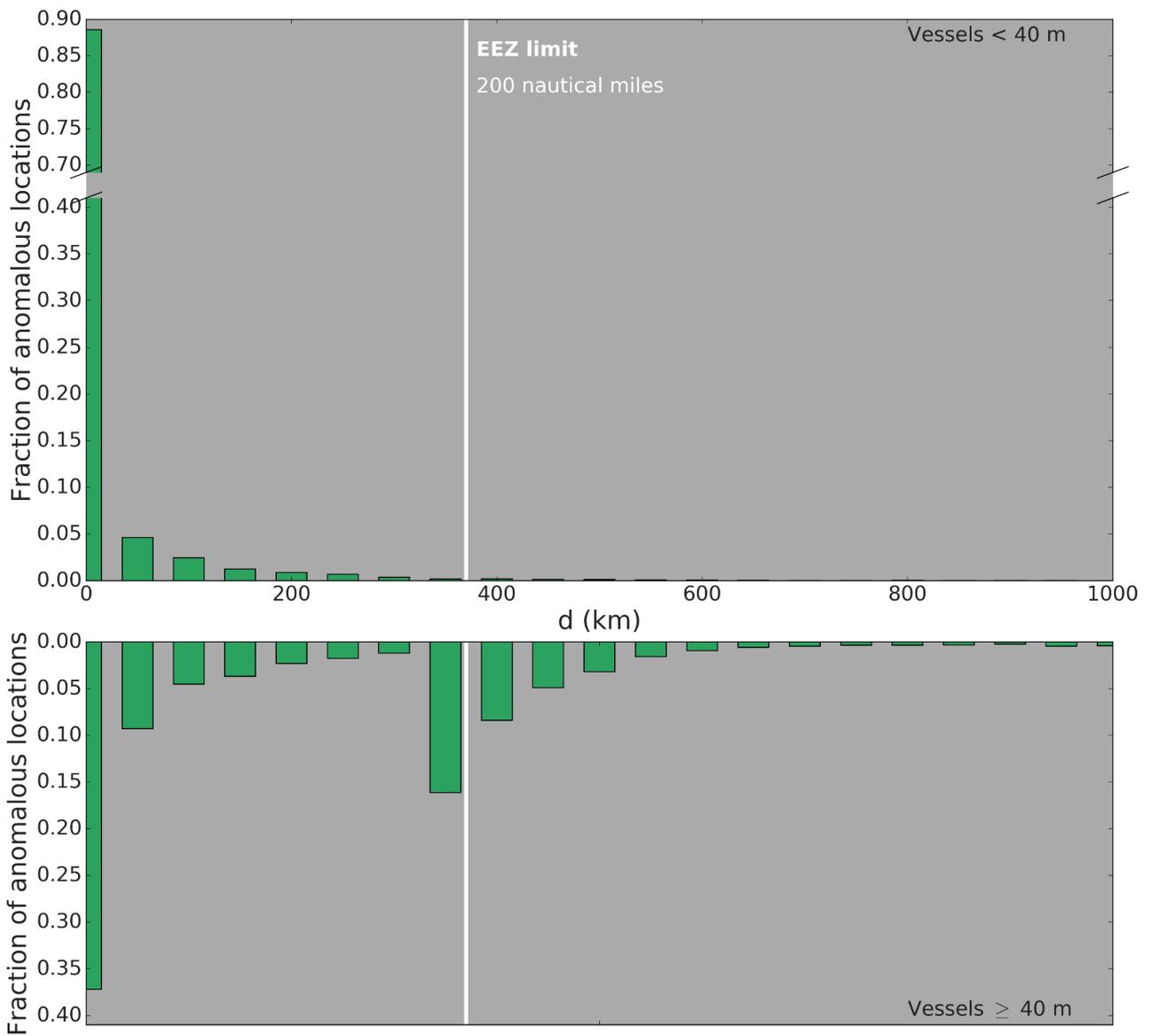

**Fig. S9.** Fraction of the global number of significant anomalies observed in cells located at a distance d from the shore, for short (top, length < 40 m) and long (bottom, length ≥40 m) fishing vessels.